\documentclass[twocolumn,showpacs,preprintnumbers,amsmath,amssymb]{revtex4}
\usepackage{graphicx}% Include figure files
\usepackage{dcolumn}% Align table columns on decimal point
\usepackage{bm}% bold math
\usepackage{amssymb}
\usepackage{epsfig}
\usepackage{rotate}
\usepackage{lscape}

\begin{document}

\title{Robustness of the helical edge states in topological insulators}
\author{Xue-Feng Wang}
%\email{xf_wang1969@yahoo.com}
\affiliation{Department of physics,
Soochow University, 1 Shizi Street, Suzhou, China 215006}
\author{Yibin Hu}
\affiliation{Centre for the Physics of Materials and Department of Physics,
McGill University, Montreal, PQ Canada H3A 2T8}
\author{Hong Guo}
%\email{guo@physics.mcgill.ca}
\affiliation{Centre for the Physics of Materials and Department of Physics,
McGill University, Montreal, PQ Canada H3A 2T8}

\begin{abstract}
Topological insulators (TI) are materials having an energy band gap in the
bulk and conducting helical electronic states on the surface. The helical states are protected by time reversal symmetry thus are expected to be robust against static disorder scattering. In this work we report atomistic first principles analysis of disorder scattering in two-probe transport junctions made of three dimensional TI material Bi$_2$Se$_3$. The robustness of the device against disorder scattering is determined quantitatively. Examining many different scattering configurations, a general trend emerges on how strong the perturbing potential and how it is spatially distributed that can derail the helical states on the Bi$_2$Se$_3$ surfaces.
\end{abstract}

\pacs{73.43.-f, 72.25.Hg, 72.10.Fk, 73.20.-r}
\maketitle

The concept of topological insulator (TI) has attracted tremendous recent attention\cite{kane1,bern,hasa,qi1} due to its fundamental role in classifying
electronic structures of materials having a band gap. TI is also important
for potential applications in nanoelectronics due to its peculiar transport properties. TI has an energy gap in its bulk band structure but has metallic helical states on its surface\cite{hasa,qi1}. Importantly and interestingly, the direction of electron spin of the helical states is locked to perpendicular to the electron momentum {\bf k}: electrons moving in the positive direction of {\bf k} have their spins pointing to one direction while those moving in negative {\bf k} pointing to exactly the opposite direction. The helical states in TI is induced by spin-orbit interaction (SOI) and are protected by time-reversal symmetry\cite{kane1,ryu}, as a result electron back scattering by disorder is suppressed because the electron's spin cannot be flipped if the disorder does not break time-reversal symmetry. The robustness of the helical conducting channels against disorder scattering is a most distinct characteristic of TI and is the most important factor for practical applications of TI to nanoelectronics.

An important question therefore arises: how robust are the helical states
in real TI materials? In other words, while weak disorder is not expected to back scatter electrons in the helical states, a strong disorder may mix the helical states with the bulk states of the disordered material and prevent helical states from establishing in the first place. Understanding how electrons traverse disordered region is urgently needed for establishing a physical picture about TI based nanoelectronics. This is especially important in light of the recent experiments in two-probe TI transport junctions in which microscopic transport parameters were measured\cite{xiu}. To the best of our knowledge, we are not aware of previous analysis on the robustness of helical states from atomic first principles that fully takes the microscopic electronic structures of the real material into account. Here, by examining many different scattering configurations \emph{ab initio}, a general trend is discovered on how strong a perturbing potential and how it is spatially distributed that can derail the helical states.

\begin{figure}
\vspace{-1 cm}
\hspace{-1.5 cm}
\includegraphics*[height=120mm,width=90mm]{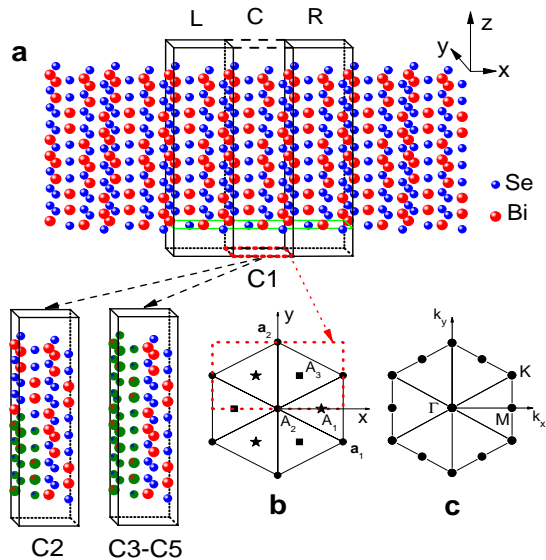}
%{BiSe_TwoProbe_6PC_NP.eps}
\vspace{-3.5 cm}
\caption{(color online)
(a)
    Atomic structure (one period along $y$-axis) of the two-probe Bi$_2$Se$_3$ transport junction. Boxes labeled L, C, and R form the two-probe unit cell composed of the left electrode, scattering region, and right electrode. Atoms to the left of L and right of R form device electrodes which extend to $x=\pm \infty$. The thickness of the Bi$_2$Se$_3$ 2D film is 6 quintuples terminating with Se atom. Each of the L, C, R boxes has two primitive cells (PC). The framed area in green between the first and second atomic layer (from bottom) is the area where electron wave function is shown in Fig.\ref{fig3}. The disorder configurations are given in Table \ref{tab1}. Region C for configurations C2, C3-C5 is shown in lower left panels where $\delta V$ is added to the atoms in green.
(b)
    Top view of the lattice vectors\cite{parameters1} $\bm{a}_1$ and $\bm{a}_2$ and the atom positions A$_1$, A$_2$, and A$_3$. The dotted rectangle shows the size of each region in the $x$-$y$ plane.
(c)
    The Brillouin zone in the $k_x$-$k_y$ plane at $k_z=0$ and the special $\bm{k}$ points $\Gamma$, M, and K.}
\label{fig1}
\end{figure}

To be specific, we consider the popular three-dimensional (3D) TI material Bi$_2$Se$_3$ forming a two-probe transport junction in the form of a two-dimensional (2D) film as shown in Fig.\ref{fig1}. Transport is in the x-direction, the film is periodic in the y-direction, and the thickness of the film (z-direction) is six quintuples or 30 atomic layers (AL) which is thick enough to isolate interactions between the top and bottom surfaces across the film\cite{zhao,zhan1,yazy}. In momentum space (k-space), the x-direction corresponds to the momentum connecting the $\Gamma$-point to the M-point (Fig.\ref{fig1}c). The two-probe device consists of three parts: left/right electrodes and the scattering region. The scattering region (indicated by letter ``C") has two primitive cells (PC) along the x-direction, a total of 60 atoms\cite{parameters1}. Potential perturbation $\delta V$ mimicking disorder is added in the scattering region of the device. Both left/right (L/R) electrodes are perfect Bi$_2$Se$_3$ films extending to $x=\pm\infty$. If there is no $\delta V$, the two-probe device reduces to a perfect infinitely big Bi$_2$Se$_3$ film.
With $\delta V$, we investigate the robustness of the incoming helical state as it traverses through the scattering region.

Our quantum transport calculation is based on first principles where density functional theory (DFT) is carried out within the nonequilibrium Green's function (NEGF) formalism\cite{tayl1}, as implemented in the Nanodcal transport package\cite{nano,foot1}. Very briefly, in the NEGF-DFT technique, the Hamiltonian of the open device structure is determined by DFT, the density matrix which includes nonequilibrium quantum statistical information of quantum transport is determined by NEGF, the transport and electro-static boundary conditions of the device are handled by real space numerical techniques. For further details we refer interested readers to the original literature\cite{tayl1}.

After the two-probe device Hamiltonian of a perfect junction (Fig.\ref{fig1}a) is converged, we analyze disorder scattering using a diagonal disorder model where a perturbing potential $\delta V$ is added to the diagonal elements of the Hamiltonian matrix corresponding to the scattering region (region C in Fig.\ref{fig1}a). This approach is adequate for our purpose while avoiding the prohibitively large computation if $\delta V$ were to be treated self-consistently.
%Recently, transport lifetime of Bi$_2$Te$_3$ nano-ribbon was measured\cite{xiu} to be $\tau_t \sim 4\times 10^{-13}$s. Assuming state life time $\tau_s\approx \tau_t$, a scattering potential is estimated to be $\delta V = h/\tau_s \sim 0.01$eV. If scattering is due to some long range potential (e.g. ionic), usually $\tau_t >> \tau_s$ and the resulting $\delta V$ can be much larger. This way,
In the study, we investigate many $\delta V$ configurations as listed in Table-\ref{tab1}.

\begin{table}
\begin{tabular}{|c|c|c|}\hline
Label & $\delta V (eV) $ & Configuration \\
\hline
C1 & 0       & perfect junction\\ \hline
C2 & 0.01    & 15 atoms, lower half PC (1-15 ALs)\\ \hline
C3 & 0.01    & 30 atoms in PC (1-30 ALs)\\ \hline
C4 & 0.1     & 30 atoms in PC (1-30 ALs)\\ \hline %%
C5 & 1.0     & 30 atoms in PC (1-30 ALs)\\ \hline %%
C6 & [0,0.1] & 60 atoms in C (1-30 ALs)\\ \hline
C7 & [0,0.5] & 60 atoms in C (1-30 ALs)\\ \hline
C8 & [0,U]   & 60 atoms in C (1-30 ALs)\\ \hline
C9 & [0,1]   & atoms in any single quintuple \\ \hline
\end{tabular}
\caption[Table-1]{Disorder configurations investigated. Configuration means where the diagonal perturbing potential $\delta V$ is added. In C2 to C5, the same value of $\delta V$ is used for all atoms specified in the configuration. PC is the first primitive cell in region C (C has two primitive cells, Fig.\ref{fig1}a). In C6 to C9, the value of $\delta V$ applied independently for each atom is drawn randomly with equal probability from the energy ranges shown.}
\label{tab1}
\end{table}

\begin{figure}
\vspace*{-2.5cm}
\hspace*{-0.3cm}
\includegraphics*[height=130mm, width=90mm]{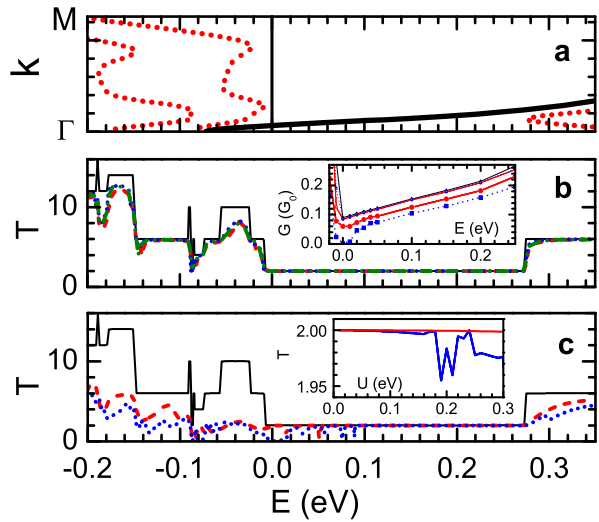}
%BiSe_TT_2PC_k00_NP.eps}
\vspace{-3cm}
\caption{(color online)
(a)
    Energy bands of the helical states (solid) and the bulk states (dotted) in the Bi$_2$Se$_3$ film between the $\bm{k}$ points M and $\Gamma$. The Fermi level is indicated by the vertical line.
(b)
    Transmission $T(E,0)$ versus electron energy $E$ near the Fermi energy $E_F$ for electrons transporting along the $\Gamma$-M direction in configuration C1 (thin solid), C3 (dashed), C4 (dotted), and C6 (dash-dotted).
    Inset: 2D conductance $G$ versus the electron energy in unit of the conductance quanta $G_o\equiv \frac{e^2}{h}$ in configurations
    C1, C4, C6, C7, and C5. (lines from up to down).
(c)
    $T(E,0)$ versus $E$ for strong disorder configurations C5 (dotted) and C7 (dashed).
    Inset: transmission at Fermi energy E=0 (dotted, blue) and E=0.1 eV (solid, red), versus the strength $U$ of random potential (C8).}
\label{fig2}
\end{figure}

Fig.\ref{fig2}a plots the calculated band structure of the C1 configuration (no disorder) - only the energy along the $\Gamma$-M direction (see Fig.\ref{fig1}c) is shown. The dotted lines are bulk bands and the solid line is the Dirac band of the helical states of the surfaces. The calculated Fermi level (vertical line) locates at round 0.072eV above the Dirac point. A gap of $\sim 0.27$eV separates the bulk valence and conduction bands while the Fermi level cuts the Dirac band. These agree with previous \emph{ab initio} results\cite{zhao}. For C1, the Fermi wave vector is calculated to be 0.0333/\AA (see Fig.\ref{fig2}a). The measurements on Bi$_2$Te$_3$ gave the same value\cite{xiu}.

For the 2D film, the transmission coefficient $T(E,k_y)$ is a function of the incoming electron energy $E$ and its transverse momentum $k_y$, and a 2D transmission $T(E)$ is obtained by integrating $T(E,k_y)$ over $k_y$. We shall first investigate $T(E,0)$ which is the transmission for electrons incoming along the $\Gamma$-M direction ($k_y=0$). We found that $T(E,0)$ has an even-integer step structure shown by the solid line in Fig.\ref{fig2}b,c. This is because there is no scattering in C1 and every incoming Bloch wave (at energy $E$) perfectly transmits through, hence $T(E,0)$ must be an even integer (even due to spin degeneracy). In particular, between the bulk gap, $T(E,0)=2$ for all energy $E$ because there is only the doubly degenerate helical state in that energy range.

Extraordinarily, $T(E,0)$ is still $2$ in the bulk gap even if there is a moderate disorder in the scattering region: as shown for C3 (dashed), C4 (dotted), and C6 (dash-dotted) configurations in Fig.\ref{fig2}b. The results of C2 is very close to that of C3 thus not plotted. These indicate that $\delta V$ cannot cause back scattering to the incoming helical states and transport is immune to it. In contrast, outside the bulk gap, i.e. for $E < -0.009$eV or $E>0.27$eV where bulk bands contribute to transport, $\delta V$ reduces the total transmission. Most surprisingly, even for the strong scattering C5 (dotted in Fig.\ref{fig2}c), $T(E,0)=2$ for the range $E=[0.1,\ 0.27]$eV. Here, $\delta V$ does reduce transmission at lower energies $E < 0.1$eV, indicating that a strong perturbing potential can prevent the surface helical state from establishing.
Our calculation also shows that the helical states are stable against a $\delta V$ as strong as 1eV if it is added to all atoms in any single quintuple of a PC in region C (configuration C9). However, a $\delta V=0.3$eV added to all atoms in two quintuples of a PC in region C might collapse the helical states. This indicates that having disorder to a larger volume is more detrimental to the helical states than increasing the disorder strength in a smaller volume. For strong random potentials of C7, the transmission (dashed line in Fig.\ref{fig2}c) is only slightly reduced from $2$ at lower energy but the helical state is destroyed near the Fermi energy. These clearly demonstrate the robustness of the helical states against disorder scattering.

An important character of helical states for perfect TI (C1) is its spin-momentum locking, namely the deviation angle $\alpha$ of the electron spin $\bm{\sigma}$ from perpendicular to its direction of momentum $\bm{e}_{\bf k}$ vanishes to a high precision\cite{zhao}: $\alpha \equiv \bm{\sigma} \cdot \bm{e}_{\bf k} =0$. In this work, we compute $\bm{\sigma}$ by spatially averaging spin in the L-C-R region (see Fig.\ref{fig1}) using the method described in Ref.[\cite{zhao}], and fix
{\bf k} along the transport direction x. With disorder, we found the angle $\alpha$ may become nonzero. In C5 and C7 where the helical states are destroyed by strong disorder, $\alpha$ is found to be as large as 0.1, ten times greater than that of other configurations.

To gain further insight, let's consider wave functions of the helical states after they suffer the $\delta V$ scattering. In perfect Bi$_2$Se$_3$ (C1), there are various Bloch states and also localized evanescent states at any given energy $E$. In the bulk gap, the Bloch states are two degenerate helical states that live on the bottom (wave function denoted as WF1) and top (WF2) surfaces, respectively, and WF2 has opposite average spin compared to WF1. During transport, the incoming helical states must traverse the scattering region (see Fig.\ref{fig1}a). In disorder configurations (C2-C9), incoming electrons in a Bloch state of spin $\sigma$ in the $n$th channel $\phi_l^n \equiv\phi_l^{n\sigma}$ (subscript $l$ indicates left incoming), are scattered by $\delta V$. The scattering process couple the incoming wave to the evanescent states and other propagating channels, the electrons are reflected or transmitted to the outgoing modes of index $m$, i.e. $\varphi_l^{m}$ or $\varphi_r^{m}$, respectively. The scattering wave function $\Psi^n_k$ corresponding to the $n$-th incoming channel is calculated as\cite{tayl1}:
\begin{equation}
\Psi^n_k=%\sum_ic^n_i\Phi_i=
\left\{
\begin{array}{ll}
\phi_l^n+\sum_m
r_m^n\varphi_l^m & {\text in\  L} \\
\psi_c^n & {\text in\  C}\\
\sum_m t_m^n\varphi_r^m &{\text in\  R}
\end{array}
\right.
\label{eq1}
\end{equation}
where $\Psi^n_k$ is expanded in terms of the extended/evanescent modes $m$. After the Hamiltonian of the two-probe device is converged by the NEGF-DFT self-consistent calculation, it is acted on $\Psi^n_k$ for a given energy $E$, and the resulting algebraic equations for the transmission ($t_m^n$) and reflection ($r_m^n$) amplitudes are calculated so that $\Psi^n_k$ can be determined\cite{tayl1}.

%Fig. 3
\begin{figure}
\vspace*{-2.5cm}
\hspace*{-0.3cm}
\includegraphics*[height=110mm, width=90mm]{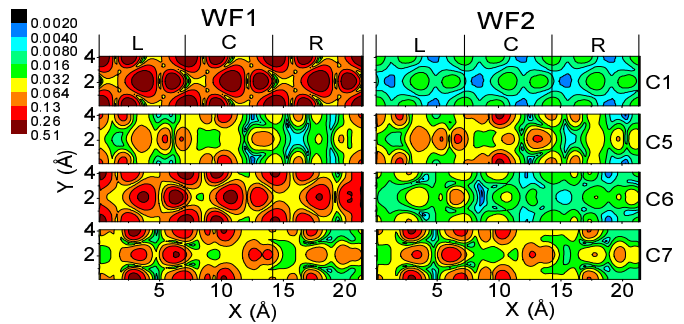}
\vspace{-4cm}
\caption{(color online) The real-space scattering states WF1 (left panels) and WF2 (right panels) in regions L, C, and R of the device for configurations C1, C5, C6, and C7 (from upper to lower panels).}
\label{fig3}
\end{figure}

Fig.\ref{fig3} plots the real space WF1 (left panels) and WF2 (right panels) sliced in the $x$-$y$ plane between the first and second ALs near the bottom surface (green frame in Fig.\ref{fig1}a). The electrons are concentrated around A$_1$ (Se atom on the first AL) and A$_2$ (Bi atom on the second AL) positions. The first row shows results for the perfect film C1. On average WF1 has an amplitude two orders of magnitude larger than WF2 since WF1 (WF2) is confined to the bottom (top) surface. In C2 which has relatively weak disorder on the 15 atoms in the lower half of the PC, WF1 is affected but WF2 remains the same as that of the C1 - because WF2 lives on the top surface away from the atoms perturbed by $\delta V$. For C3 and C4 where $\delta V$ is added across the film, both WF1 and WF2 are significantly affected in region C but much less so out side it, hence the transmission and the 2D conductance is not affected by $\delta V$ (Fig.\ref{fig2}). The patterns of real space WF1 and WF2 for C2-C4 are similar to those of C1 (not shown). For C5 (second row  in Fig.\ref{fig3}) which is a strong $\delta V$ case, WF1 and WF2 are no longer separated in real space and have similar orbital distributions with opposite spin orientations. The two states are now mixed through the film with almost symmetric profiles of amplitude along $z$ between the two surfaces. Namely, scattering states in regions L, C, R are all significantly affected thus transmission reduces (indicated in Fig.\ref{fig2}). Our calculation indicates that a $\delta V$ of $\sim 0.3$eV destroys the helical states if being applied to a region enclosing more than $\sim 10$ atoms. If $\delta V$ is added to any single atom - a single impurity, the helical states are robust up to $\delta V \approx 1$eV which is a very large local perturbation indeed. Since there are 60 atoms in region C, 10 perturbed atoms correspond to an impurity concentration of 16.7\%, while a single impurity corresponds to a concentration of 1.67\%.

Similar robustness of the helical states is found for random potentials: for C6 (third row), both WF1 and WF2 are modified from the perfect film case but they are still localized in their respective surfaces thus transmission $T=2$ in the bulk gap is largely intact. For C6 the \emph{average} perturbing potential $\delta V=0.05$eV, not strong enough to derail the helical states. For C7 - stronger random potential (fourth row), WF1 and WF2 become similar in real space (i.e. not confined in surfaces) indicating the helical states are destroyed. It is very interesting to note that even though $\delta V$ is added only in region C, the scattering states (WF1, WF2) are affected in all regions L, C, and R reflecting the scattering processes.

To correlate the robustness of transmission $T(E,0)$ against the strength of the disorder potential $\delta V$, we have calculated configuration C8 (see Table-\ref{tab1}) where $\delta V$ is randomly drawn from ranges $[0,U]$, results are shown in the inset of Fig.\ref{fig2}c versus $U$. At Fermi energy $E=0$ (blue dotted line), $T(E,0) = 2$ until $U$ reaches $\sim 0.2$eV at which $T(E,0)$ substantially reduces (inset of Fig.\ref{fig2}c). By investigating which atomic orbitals contribute to the scattering wave functions and also the real space WF1/WF2, the scattering states are found to be unconfined from the surfaces at this value of $U$, namely the helical states are destroyed. When $U$ increases further, $T(E,0)$ oscillates by the random disorder scattering configurations. At higher energies, we found that the helical edge states are much more robust. For instance, at $E=0.1$eV (solid curve in the inset of Fig.\ref{fig2}c), $T(E,0)\approx 2$ until $U \sim 0.5$eV. This is reasonable since electrons with higher energy are less affected by potential perturbations.

So far we focused on transmission in the x-direction, $T(E,k_y=0)$. For the film,
a 2D transmission $T(E)$ is obtained by integrating $T(E,k_y)$ over $k_y$, which gives the 2D conductance $G(E) = T(E)\times G_o$ where $G_o\equiv e^2/h$. The inset of Fig.\ref{fig2}b plots $G(E)$. For energies outside the bulk gap, e.g. $E < 0$ or $E \gtrsim 0.27$eV, bulk propagating modes participate transport and therefore $G$ decreases according to the increase of the disorder. Within the gap, $G(E)$ changes negligibly for weak disorder C4 and C6, but more significantly in strong disorder C5 and C7. In contrast to $T(E,0)$ (the main figure in Fig.\ref{fig2}b,c), $G(E)$ shows a significant decrease throughout the gap for strong disorder C5 and C7.

In summary, a general trend emerges from the {\it ab initio} two-probe device simulation concerning the robustness of the surface helical states of TI material Bi$_2$Se$_3$. For the Bi$_2$Se$_3$ 2D film with six quintuple thickness, a perturbing potential of $\sim 0.1$eV or less in the scattering region of one primitive cell wide, is difficult to destroy the helical states. Larger perturbing potential may destroy the helical states if applied to regions containing roughly 10 atoms in each unit cell. A single impurity in each unit cell - corresponding to 1.67\% density, does not derail helical states up to $\delta V \approx 1$eV. The helical states at higher energies are much harder to destroy by the disorder potential. When helical states are destroyed by disorder, they are no longer localized on the surfaces but are spatially mixed throughout the film. Microscopically, when propagating helical states pass through a disordered region, evanescent states are excited around the perturbed atoms and the scattering states are modified.

{\bf Acknowledgement.}
We are grateful to Dr. Lei Liu and Eric Zhu for assistance in numerical calculations using the nanodcal software, and to Prof. Y.H. Zhao for sharing the TZDP basis functions. X.F.W. thanks Drs. Jingzhe Chen and Xun Xue for helpful discussions. X.F.W is supported by NSFC of China (Nos. 11074182 and 91121021). H.G. is supported by NSERC of Canada, FQRNT of Quebec and CIFAR. We gratefully acknowledge RQCHP and CLUMEQ for computing facilities.


\begin{thebibliography}{99}
\bibitem{parameters1}
    A vacuum of thickness 20 \AA\ is added on top of the Bi$_2$Se$_3$ film to form the calculation cell that is repeated in the x-z directions. The lattice vector of the primitive cell (PC) is $a_1=(\sqrt{3}a/2,-a/2,0)$, $a_2=(0,a,0)$, $a_3=(0,0,2c+15.9\AA)$ where $a=4.107$ \AA\ and $c=27.89$ \AA.
\end{thebibliography}
\end{document}